\documentclass[3p,times]{elsarticle}

\usepackage{ecrc}

\volume{00}

\firstpage{1}

\journalname{Nuclear Physics A}

\runauth{B. S. Zou}

\jid{NUPHA}

\jnltitlelogo{Nuclear Physics A}

\CopyrightLine{2012}{Published by Elsevier Ltd.}

\usepackage{amssymb}

\usepackage[figuresright]{rotating}

\begin{document}

\begin{frontmatter}

\dochead{}



\title{Hadron spectroscopy from strangeness to charm and beauty}

\author{B.~S.~Zou}

\address{1) State Key Laboratory of Theoretical Physics, Institute of
Theoretical Physics, Chinese Academy of Sciences, Beijing 100190,
China\\2) Institute of High Energy Physics and Theoretical Physics
Center for Science Facilities, Chinese Academy of Sciences, Beijing
100049, China}


\begin{abstract}
Quarks of different flavors have different masses, which will cause
breaking of flavor symmetries of QCD. Flavor symmetries and their
breaking in hadron spectroscopy play important role for
understanding the internal structures of hadrons. Hadron
spectroscopy with strangeness reveals the importance of unquenched
quark dynamics. Systematic study of hadron spectroscopy with
strange, charm and beauty quarks would be very revealing and
essential for understanding the internal structure of hadrons and
its underlying quark dynamics.
\end{abstract}


\end{frontmatter}


\section{Hadron spectroscopy with strangeness}
\label{}

In classical constituent quark models, baryons are ascribed as
three-quark ($qqq$) states and mesons are ascribed as
quark-anti-quark ($q\bar q$) states. This picture is very successful
in explaining properties of the spatial ground states of the flavor
SU(3) vector meson nonet, baryon octet and decuplet. Its predicted
$\Omega(sss)$ baryon with mass around 1670 MeV was discovered by
later experiments. However even for the lowest spatial excited
states, this picture failed badly in both meson and baryon sectors.

In the meson sector, the lowest spatial excited SU(3) nonet is the
scalar nonet composed of $f_0(500)$, $\kappa(600\sim 800)$,
$a_0(980)$ and $f_0(980)$. In the classical constituent quark
models, these scalars should be $q\bar q~(L=1)$ states with
$f_0(500)$ as $(u\bar u+d\bar d)/\sqrt{2}$ state, $a_0^0(980)$ as
$(u\bar u-d\bar d)/\sqrt{2}$ state and $f_0(980)$ as mainly $s\bar
s$ state. Then in this picture, it cannot explain why the mass of
$a_0(980)$ is degenerate with $f_0(980)$ instead of close to
$f_0(500)$ as in the $\rho$-$\omega$ case in the vector nonet. This
made R.J.Jaffe~\cite{Jaffe77} proposing these scalars are $q^2\bar
q^2$ states instead of $q\bar q$ states. In the new picture, the
$f_0(500)$ is ascribed as $[ud]\overline{[ud]}$ state, $a_0^0(980)$
as $([us]\overline{[us]}-[ds]\overline{[ds]})/\sqrt{2}$ state and
$f_0(980)$ as $([us]\overline{[us]}+[ds]\overline{[ds]})/\sqrt{2}$
state. This gives a natural explanation of the degeneracy of
$a_0(980)$ and $f_0(980)$. Here $[q_1q_2]$ means a good diquark with
configuration of flavor representation  ${\bf\bar 3}$, spin 0 and
color $\bar 3$. Alternatively, these scalars are also proposed to be
meson-meson dynamically generated states~\cite{Isgur82,Oller}.

In the baryon sector, the similar thing seems also
happening~\cite{zou2008}.  In the classical quark models, the
excited baryon states are described as excitation of individual
constituent quarks, similar to the cases for atomic and nuclear
excitations. The lowest spatial excited baryon is expected to be a
($uud$) $N^*$ state with one quark in orbital angular momentum $L=1$
state, and spin-parity $1/2^-$. However, experimentally, the lowest
negative parity $N^*$ resonance is found to be $N^*(1535)$, which is
heavier than two other spatial excited baryons: $\Lambda^*(1405)$
and $N^*(1440)$. This is the long-standing mass reverse problem for
the lowest spatial excited baryons.

In the simple 3q constituent quark models, it is also difficult to
understand the strange decay properties of the $N^*(1535)$, which
seems to couple strongly to the final states with strangeness.
Besides a large coupling to $N\eta$, a large value of
$g_{N^*(1535)K\Lambda}$ is deduced~\cite{liubc,gengls} by a
simultaneous fit to BES data on $J/\psi\to\bar pp\eta$,
$pK^-\bar\Lambda+c.c.$, and COSY data on $pp\to pK^+\Lambda$. There
is also evidence for large $g_{N^*(1535)N\eta^\prime}$ coupling from
$\gamma p \to p\eta^\prime$ reaction at CLAS~\cite{etap} and $pp\to
pp\eta^\prime$ reaction~\cite{caox}, and large $g_{N^*(1535)N\phi}$
coupling from $\pi^- p \to n\phi$, $pp\to pp\phi$ and $pn\to d\phi$
reactions~\cite{xiejj1,Doring,caox2009}.

The third difficulty is the strange decay pattern of another member
of the $1/2^-$-nonet, $\Lambda^*(1670)$, which has its coupling to
$\Lambda\eta$ much larger than $NK$ and $\Sigma\pi$ according to its
branching ratios listed in PDG~\cite{PDG}.

All these difficulties can be easily understood by considering large
5-quark components in them~\cite{zou2008,liubc,Helminen,zhusl}. The
$N^*(1535)$ could be the lowest $L=1$ orbital excited $|uud>$ state
with a large admixture of $|[ud][us]\bar s>$ pentaquark component
having $[ud]$, $[us]$ and $\bar s$ in the ground state. The
$N^*(1440)$ could be the lowest radial excited $|uud>$ state with a
large admixture of $|[ud][ud]\bar d>$ pentaquark component having
two $[ud]$ diquarks in the relative P-wave. While the lowest $L=1$
orbital excited $|uud>$ state should have a mass lower than the
lowest radial excited $|uud>$ state, the $|[ud][us]\bar s>$
pentaquark component has a higher mass than $|[ud][ud]\bar d>$
pentaquark component. The lighter $\Lambda^*(1405)1/2^-$ is also
understandable in this picture. Its main 5-quark configuration is
$|[ud][qs]\bar q>$ which is lighter than the corresponding 5-quark
configuration $|[ud][us]\bar s>$ in the $N^*(1535)1/2^-$. The large
mixture of the $|[ud][us]\bar s>$ pentaquark component in the
$N^*(1535)$ naturally results in its large couplings to the $N\eta$,
$N\eta^\prime$, $N\phi$ and $K\Lambda$. The main 5-quark
configuration for the $\Lambda^*(1670)$ is $|[us][ds]\bar s>$ which
makes it heavier than other $1/2^-$ states and larger coupling to
$\Lambda\eta$.

Besides the penta-quark configurations with the diquark correlation,
the penta-quark system may also be in the form of meson-baryon
states. The $N^*(1535)$, $\Lambda^*(1405)$ and some other baryon
resonances are proposed to be meson-baryon dynamically generated
states~\cite{Oller,Weise,or,meiss,Inoue,lutz,Hyodopk}. However, a
challenge for this meson-baryon dynamical picture is to explain the
mass and decay pattern of the $\Lambda^*(1670)$.

From above facts and discussion for both meson and baryon sectors,
one can see that unlike atomic and nuclear excitations where the
number of constituent particles are fixed, the favorable hadronic
excitation mechanism for the lowest spatial excited states in light
quark sector seems to be dragging out a light $q\bar q$ pair from
gluon field rather than to excite a constituent quark to be $L=1$
state. A breathing mode of $qqq\leftrightarrow qqqq\bar q$ is
proposed~\cite{an2009,zouHypX} for the lowest $1/2^-$ baryon nonet.
Each baryon is a mixture of the three-quark and five-quark
components.

While the new picture gives a nice account of properties of scalar
meson nonet and well-established members of the lowest $1/2^-$
baryon nonet, it is necessary to check its distinguishable
predictions of other members in the $1/2^-$ baryon nonet. While the
classical quenched quark models~\cite{Capstick} predict the $1/2^-$
$\Sigma^*$ and $\Xi^*$ to be around 1650 MeV and 1760 MeV,
respectively, the unquenched quark
models~\cite{Helminen,zhusl,zouHypX} expect them to be around 1400
MeV and 1550 MeV, respectively, and meson-baryon dynamical
models~\cite{Oh,Kanchan,Ramos} predict them to be around 1450 MeV
and 1620 MeV, respectively.

For $\Sigma$ resonance with $J^P={1\over 2}^-$, the PDG~\cite{PDG}
lists a two-star $ \Sigma(1620)$ resonance, which seems to support
the quenched quark models. However, only four references listed in
PDG show weak evidence for its existence. In Ref.~\cite{16202}, the
total cross sections for $K^-p$ and $K^-n$ are analyzed, which
indicates some $\Sigma$ resonance around 1600 MeV without $J^P$
quantum number. Refs.~\cite{16208,16207} are based on multichannel
analysis of the $\overline{K}N$ reactions. Both claim evidence for a
$\Sigma~{1\over 2}^-$ resonance with mass around 1620 MeV but give
contradicted coupling properties to $\pi\Lambda$ and to $\pi\Sigma$.
Other later multichannel analyses of the $\overline{K}N$ reactions
support the existence of an $\Sigma(1660){1\over 2}^+$~\cite{PDG}.
Ref.~\cite{16201} analyzes the reaction
$K^-n\rightarrow\pi^-\Lambda$ and gives two comparable solutions
with and without $\Sigma(1620){1\over 2}^-$.

On the other hand, there are also some supports of the unquenched
5-quark models with $\Sigma^*({1\over 2}^-)$ of much lower masses.
The re-analysis of old data on $K^-p\to\Lambda\pi^+\pi^-$ finds
hidden $\Sigma^*({1\over 2}^-)$ with mass around 1380 MeV under the
$\Sigma(1385){3\over 2}^+$ peak~\cite{Wujj}. From an analysis of the
recent LEPS data on $\gamma n\to K^+\Sigma^{*-}(1385)$~\cite{LEPS},
there is also a possibility for the existence of such low mass
$\Sigma^*({1\over 2}^-)$~\cite{pu1380}. An analysis of CEBAF data on
$\gamma p\to K^+\pi\Sigma$ also suggests a possible
$\Sigma^*({1\over 2}^-)$ around 1400 MeV~\cite{Schumacher}.

To clarify the situation for $\Sigma$ resonances, recently, a
combined fit for the new CB data~\cite{CryB} on
$K^-p\to\pi^0\Lambda$ together with the old data~\cite{16201} on
$K^-n\to\pi^-\Lambda$ for the energies from 1569 to 1676 MeV was
performed~\cite{shi12}. The $\overline{K}N\rightarrow\pi\Lambda$
reaction is the best channel available for the study of the $\Sigma$
resonances because the $\pi\Lambda$ is a pure isospin 1 channel. The
high precision Crystal Ball $\Lambda$ polarization data~\cite{CryB}
are crucial for discriminating $\Sigma(1620){1\over 2}^-$ from
$\Sigma(1635){1\over 2}^+$. It shows that the $\Sigma(1660){1\over
2}^+$ is definitely needed, while $\Sigma(1620){1\over 2}^-$ is not
needed at all. Although $\Sigma(1380){1\over 2}^-$ is not demanded
in this analysis, it cannot be excluded. Therefore, no evidence to
support the classical quenched quark models at all from the $1/2^-$
baryons. Additional $\Sigma(1542){3\over 2}^-$, $\Sigma(1840){3\over
2}^+$ and $\Sigma(1610){1\over 2}^+$ may exist.

The $\Sigma(1542){3\over 2}^-$ is consistent with the resonance
structure $\Sigma(1560)$ or $\Sigma(1580){3\over 2}^-$ in
PDG~\cite{PDG} and seems a good isospin partner of
$\Lambda(1520){3\over 2}^-$. Recently a very interesting narrow
$\Lambda(1670){3\over 2}^-$ with a width about 1.5 MeV was claimed
from an analysis of $K^-p\to\eta\Lambda$ data~\cite{Liubc2012}.
Together with $N^*(1520){3\over 2}^-$ and either $\Xi(1620)$ or
$\Xi(1690)$, they fit in a nice $3/2^-$ baryon nonet with large
penta-quark configuration, {\sl i.e.}, $N^*(1520)$ as
$|[ud]\{uq\}\bar q>$ state, $\Lambda(1520)$ as $|[ud]\{sq\}\bar q>$
state, $\Lambda(1670)$ as $|[ud]\{ss\}\bar s>$ state, and
$\Xi(16xx)$ as $|[ud]\{ss\}\bar q>$ state. Here $\{q_1q_2\}$ means a
diquark with configuration of flavor representation ${\bf 6}$, spin
1 and color $\bar 3$. The $\Lambda(1670)$ as $|[ud]\{ss\}\bar s>$
state gives a natural explanation for its dominant $\eta\Lambda$
decay mode with a very narrow width due to its very small phase
space meanwhile a D-wave decay.

The available information on the hadron spectroscopy with
strangeness strongly indicates that $qqqq\bar q$ in S-state is more
favorable than $qqq$ state with $L=1$ and $q^2\bar q^2$ in S-state
is more favorable than $q\bar q$ state with $L=1$. The multi-quark
components are very substantial and important for hadronic excited
states. Even $q^6\bar q$ configuration may play dominant role for
some baryon resonances~\cite{Gal,Oset2008}.

To further establish the multi-quark picture for hadronic excited
states, it is very important to complete the low-lying hyperon
spectrum, especially the $1/2^-$ and $3/2^-$ $\Sigma^*$, $\Xi^*$ and
$\Omega^*$. Here the $\Omega^*$ spectrum has a unique advantage that
the favorable $q\bar q$ excitations from quark sea have different
flavor from the valence strange quarks~\cite{An2012}. Kaon beam
experiment at JPARC and hyperon production from charmonium decays at
BESIII may play very important role in this aspect. It is also
important to check the cases with $s$ quarks replaced by $c$ or $b$
quarks.

\section{From strangeness to charm and beauty}

Various pictures and dynamics for the spectroscopy with strangeness
can be extended to and checked by its charm and beauty partners. For
example, if $f_0(980)$ is a $K\bar K$ molecule mainly due to light
vector meson exchange force~\cite{Krewald2003}, then with the same
mechanism there should also exist $DK$, $B\bar K$, $D\bar D$ and
$\bar BB$ molecules~\cite{Zhangyj,Guofk2006}. The newly established
$D^*_{s0}(2317)$ is regarded as a $DK$ molecule or tetra-quark state
by many people~\cite{Klempt2007}. The $f_1(1420)$ was proposed to be
a $K^*\bar K$ molecule~\cite{Tornqvist}; now the newly established
$X(3872)$ is regarded as its $D^*\bar D$
partner~\cite{Zhusl09,Vijande}. The $\Lambda_c(2595)1/2^-$ was
proposed~\cite{Tolos} to be $DN$ molecule as the charm partner of
$\Lambda(1405)$.

Although many hadron resonances were proposed to be hadron-hadron
dynamically generated states or multi-quark states, most of them
cannot be clearly distinguished from classical quark model states
due to tunable ingredients and possible large mixing of various
configurations in these models. Even in 2010, the PDG~\cite{PDG2010}
still claimed that ``The clean $\Lambda_c$ spectrum has in fact been
taken to settle the decades-long discussion about the nature of the
$\Lambda(1405)$ -- true 3-quark state or mere $\bar KN$ threshold
effect? -- unambiguously in favor of the first interpretation." A
possible solution to this problem is to extend the penta-quark study
to the hidden charm and hidden beauty sectors. If the $N^*(1535)$ is
the $\bar K\Sigma$ quasi-bound state with hidden strangeness, then
naturally by replacing $s\bar s$ by $c\bar c$ or $b\bar b$ one would
expect super-heavy $N^*$ states with hidden charm and hidden beauty
just below $\bar D\Sigma_c$ and $B\Sigma_b$ thresholds,
respectively.

Following the Valencia approach of Ref.\cite{ramos} and extending it
to the hidden charm sector, the interaction between various charmed
mesons and charmed baryons were studied with the local hidden gauge
formalism in Refs.\cite{Wujj1,Wujj2}. Several meson-baryon
dynamically generated narrow $N^*$ and $\Lambda^*$ resonances with
hidden charm are predicted with mass around 4.3 GeV and width
smaller than 100 MeV. The S-wave $\Sigma_c\bar D$ and $\Lambda_c\bar
D$ states with isospin I=1/2 and spin S=1/2 were also investigated
by various other approaches~\cite{Wangwl,Wujj-Lee,Liuxiang}. They
confirm that the interaction between $\Sigma_c$ and $\bar D$ is
attractive and results in a $\Sigma_c\bar D$ bound state not far
below threshold. The low-lying energy spectra of five quark systems
$uudc\bar{c}$ ($I$=$1/2$, $S$=$0$) and $udsc\bar{c}$ ($I$=$0$,
$S$=$-1$) are also investigated with three kinds of schematic
interactions: the chromomagnetic interaction, the flavor-spin
dependent interaction and the instanton-induced
interaction~\cite{Yuansg}. In all the three models, the lowest five
quark state ($uudc\bar{c}$ or $udsc\bar{c}$) has an orbital angular
momentum $L=0$ and the spin-parity $J^{P}=1/2^{-}$; the mass of the
lowest $udsc\bar{c}$ state is heavier than the lowest $uudc\bar{c}$
state, which is different from the prediction of meson-baryon
dynamical model~\cite{Wujj1,Wujj2}. The predicted new resonances
definitely cannot be accommodated by quark models with three
constituent quarks. Because these predicted states have masses above
$\eta_cN$ and $\eta_c\Lambda$ thresholds, they can be looked for at
the forthcoming PANDA/FAIR and JLab 12-GeV upgrade experiments. This
is an advantage for their experimental searches, compared with those
baryons with hidden charms below the $\eta_cN$ threshold proposed by
other earlier approaches~\cite{Gobbi,Hofmann}.

The same meson-baryon coupled channel unitary approach with the
local hidden gauge formalism was extended to the hidden beauty
sector in Ref.\cite{Wujj3}. Two $N^*_{b\bar b}$ states and four
$\Lambda^*_{b\bar b}$ states were predicted to be dynamically
generated. Because of the hidden $b\bar{b}$ components involved in
these states, the masses of these states are all above 11 GeV while
their widths are of only a few MeV, which should form part of the
heaviest island for the quite stable $N^*$ and $\Lambda^*$ baryons.
For the Valencia approach, the static limit is assumed for the
t-channel exchange of light vector mesons by neglecting momentum
dependent terms. In order to investigate the possible influence of
the momentum dependent terms, the conventional Schrodinger Equation
approach was also used to study possible bound states for the
$B\Sigma_b$ channel by keeping the momentum dependent terms in the
t-channel meson exchange potential. It was found that within the
reasonable model parameter range the two approaches give consistent
predictions about possible bound states. This gives some
justification of the simple Valencia approach although there could
be an uncertainty of 10 - 20 MeV for the binding energies.

Production cross sections of the predicted $N^*_{b\bar{b}}$
resonances in $pp$ and $ep$ collisions were estimated as a guide for
the possible experimental search at relevant facilities in the
future. For the $pp \to pp \eta_b$ reaction, the best center-of-mass
energy for observing the predicted $N^*_{b\bar{b}}$ is $13\sim 25$
GeV, where the production cross section is about 0.01 nb. For the
$e^-p \to e^-p \Upsilon$ reaction, when the center-of-mass energy is
larger than 14 GeV, the production cross section should be larger
than 0.1 nb. Nowadays, the luminosity for pp or ep collisions can
reach $10^{33}cm^{-2}s^{-1}$, this will produce more than 1000
events per day for the $N^*_{b\bar{b}}$ production. It is expected
that future facilities, such as proposed electron-ion collider
(EIC), may discover these very interesting super-heavy $N^*$ and
$\Lambda^*$ with hidden beauty.

Very recently, the observation of the iso-vector meson partners of
the predicted $N^*_{b\bar{b}}$, $Z_b(10610)$ and $Z_b(10650)$, were
reported by Belle Collaboration~\cite{Zb}. This gives us stronger
confidence on the existence of the super-heavy island for the
$N^*_{b\bar{b}}$ and $\Lambda^*_{b\bar{b}}$ resonances.

\section{Conclusions}

Available information on hadron spectroscopy with strangeness and
charm reveals unquenched quark picture. Dragging out a $q\bar q$
from gluon field is a very important excitation mechanism for
hadrons. To correctly describe the hadron spectrum, it is necessary
to go beyond the classical quenched quark models which assuming a
fixed number of constituent quarks. Distinguishable prediction for
hyperon spectroscopy from the new picture is yelling for
experimental confirmation. Kaon beam experiments at JPARC and
hyperon production data from charmonium decays at BESIII can play
very important role here. Super-heavy narrow $N^*$ and $\Lambda^*$
resonances are predicted by various models to exist around 4.3 GeV
and 11 GeV for hidden charm and beauty, respectively. Their
iso-vector meson partners $Z_b(10610)$ and  $Z_b(10650)$ have
recently been observed. Experimental confirmation of them will
unambiguously establish multi-quark dynamics. They can be looked for
at CEBAF-12GeV-upgrade at Jlab and PANDA at FAIR, maybe also at
JPARC, super-B, RHIC , EIC.

\section{Acknowledgments}
I thank C.~S.~An, P.~Z.~Gao, J.~He, F.~Huang, T.~S.~H.~Lee,
R.~Molina, E.~Oset, J.~Shi, W.~L.~Wang, K.~W.~Wei, J.~J.~Wu,
H.~S.~Xu, S.~G.~Yuan, L.~Zhao, Z.~Y.~Zhang for collaboration works
reviewed here. This work is supported by the National Natural
Science Foundation of China under Grant 11035006, 11121092,
11261130311 (CRC110 by DFG and NSFC), the Chinese Academy of
Sciences under Project No.KJCX2-EW-N01 and the Ministry of Science
and Technology of China (2009CB825200).






\begin{thebibliography}{00}


\bibitem{Jaffe77} R. J. Jaffe,  Phys.\ Rev. \ D15 (1977) 267.

\bibitem{Isgur82} J. Weinstein and N. Isgur, Phys.\ Rev.\ Lett. 48
(1982) 659.

\bibitem{Oller}  J. A. Oller, E. Oset, A. Ramos, Prog.\ Part.\ Nucl.\ Phys. 45 (2000) 157, and
references therein.

\bibitem{zou2008} B. S. Zou, Eur.\ Phys. \ J. \ A35 (2008) 325.

\bibitem{liubc} B. C. Liu, B. S. Zou, Phys.\ Rev.\ Lett.\   96 (2006)
042002.

\bibitem{gengls} L. S. Geng, E. Oset, B. S. Zou, M. Doring, Phys.\ Rev.\  C79 (2009) 025203.

\bibitem{etap} M. Dugger et al., Phys.\ Rev.\ Lett.\  96 (2006)
062001.

\bibitem{caox} X. Cao, X. G. Lee, Phys.\ Rev.\ C78 (2008) 035207.

\bibitem{xiejj1} J. J. Xie, B. S. Zou, H.C. Chiang, Phys.\ Rev.\  C77 (2008)
015206.

\bibitem{Doring}
M. Doring, E. Oset, B. S. Zou, Phys.\ Rev.\  C78 (2008) 025207.

\bibitem{caox2009} X. Cao, J. J. Xie, B. S. Zou, H. S. Xu, Phys.\ Rev.\  C80 (2009)
025203.

\bibitem{PDG} J. Beringer et al. (Particle Data Group), Phys.\ Rev.\ D86 (2012) 010001.

\bibitem{Helminen} C. Helminen, D. O. Riska, Nucl.\ Phys.\  A699 (2002)
624.

\bibitem{zhusl} A. Zhang et al., High Energy Phys.\ Nucl.\ Phys.\  29 (2005)
250.

\bibitem{Weise} N. Kaiser, P. B. Siegel, W. Weise,
Phys.\ Lett.\  B362 (1995) 23.

\bibitem{or} E. Oset, A. Ramos, Nucl.\ Phys.\  A635 (1998) 99.

\bibitem{meiss} J. A. Oller, U. G. Meissner, Phys.\ Lett.\  B 500 (2001) 263.

\bibitem{Inoue} T. Inoue, E. Oset, M. J. Vicente Vacas, Phys.\ Rev.\  C {\bf 65}, 035204 (2002)

\bibitem{lutz}  C. Garcia-Recio, M. F. M. Lutz, J. Nieves, Phys.\ Lett.\  B582
(2004) 49.

\bibitem{Hyodopk} T. Hyodo, S. I. Nam, D. Jido, A. Hosaka, Phys.\ Rev.\  C68 (2003) 018201.

\bibitem{an2009} C. S. An, B. S. Zou, Eur.\ Phys.\ J.\  A39 (2009) 195.

\bibitem{zouHypX} B. S. Zou,  Nucl.\ Phys.\  A835 (2010) 199.

\bibitem{Capstick} S. Capstick, W. Roberts, Prog. Part. Nucl. Phys. 45 (2000) S241.

\bibitem{Oh} Y. Oh, Phys.\ Rev.\  D75 (2007) 074002.

\bibitem{Kanchan} K. P. Khemchandani, A. M. Torres, H. Kaneko, H. Nagahiro,
A. Hosaka, Phys.\ Rev.\ D {\bf 84} (2011) 094018.

\bibitem{Ramos} A. Ramos, E. Oset, C. Bennhold, Phys.\ Rev.\ Lett.\  {\bf 89} (2002)
252001.

%
\bibitem{16202} A. S. Carroll {\it et al.}, Phys.\ Rev.\ Lett. {\bf 37} (1976) 806.
%
\bibitem{16208} J. K. Kim, Phys.\ Rev.\ Lett. {\bf 27} (1971) 356.
%
\bibitem{16207} W. Langbein, F. Wagner, Nucl. Phys. B47 (1972) 477.
%
\bibitem{16201} W. A. Morris {\it et al.}, Phys.\ Rev.\ D17 (1978) 55.

\bibitem{Wujj} J. J. Wu, S. Dulat, B. S. Zou, Phys.\ Rev.\  {\bf D80} (2009) 017503;
Phys.\ Rev.\ {\bf C81} (2010) 045210.
%
\bibitem{LEPS} K. Hicks {\it et al} (LEPS Collaboration), Phys.\ Rev.\ Lett.\
{\bf102} (2009) 012501.
%
\bibitem{pu1380} P. Gao, J. J. Wu, B. S. Zou, Phys.\ Rev.\  {\bf C81} (2010) 055203.

\bibitem{Schumacher} R. Schumacher, in this proceedings.
%
\bibitem{CryB} S. Prakhov {\it et al}, Phys.\ Rev.\  {\bf C80} (2009) 025204.

\bibitem{shi12} P. Gao, J. Shi, B.S. Zou, Phys.\ Rev.\  {\bf C86} (2012) 025201.

\bibitem{Liubc2012} B.~C.~Liu, J.~J.~Xie, Phys.\ Rev.\ C {\bf 86} (2012)
055202; Phys.\ Rev.\ C {\bf 85} (2012) 038201.

%
\bibitem{Gal} A. Gal, H. Garcilazo, Nucl.\ Phys.\  {\bf A864} (2011) 153.
%
\bibitem{Oset2008} A. M. Torres, K.P. Khemchandani, E. Oset, Phys.\ Rev.\  {\bf C77} (2008) 042203.
%
\bibitem{An2012} S.~G.~Yuan, C.~S.~An, K.~W.~Wei, B.~S.~Zou, H.~S.~Xu,
arXiv:1208.1742 [hep-ph].
%
\bibitem{Krewald2003} S.~Krewald, R.~H.~Lemmer and F.~P.~Sassen,
 Phys.\ Rev.\ D {\bf 69} (2004) 016003.
%
\bibitem{Zhangyj} Y.~J.~Zhang, H.~C.~Chiang, P.~N.~Shen, B.~S.~Zou,
Phys.\ Rev.\ D {\bf 74} (2006) 014013.
%
\bibitem{Guofk2006} F.~K.~Guo, P.~N.~Shen, H.~C.~Chiang, R.~G.~Ping and B.~S.~Zou,
Phys.\ Lett.\ B {\bf 641} (2006) 278.
%
\bibitem{Klempt2007} E.~Klempt and A.~Zaitsev, Phys.\ Rept.\  {\bf 454} (2007)
1, and references therein.
%
\bibitem{Tornqvist} N.~A.~Tornqvist, Phys.\ Rev.\ Lett.\  {\bf 67} (1991) 556.
%
\bibitem{Zhusl09} X.~Liu, Z.~G.~Luo, Y.~R.~Liu and S.~L.~Zhu, Eur.\ Phys.\ J.\ C {\bf 61} (2009) 411;
Phys. Rev. D84 (2011) 054002.
%
\bibitem{Vijande} T.~Fernandez-Carames, A.~Valcarce and J.~Vijande,  Phys.\ Rev.\ Lett.\  {\bf 103} (2009)
222001.
%
\bibitem{Tolos} L. Tolos et al., Chin.\ Phys.\ C {\bf 33} (2009)
1323; J. Haidenbauer et al., Eur. Phys. J A47 (2011) 18.
%
\bibitem{PDG2010} K. Nakamura {\it et al.} (Particle Data Group), J.\ Phys.\ G {\bf 37} (2010)
075021.
%
\bibitem{ramos} E. Oset, A. Ramos, Euro.\ Phys.\ J.\ A {\bf 44}
(2010) 445.
%
\bibitem{Wujj1} J.~J.~Wu, R.~Molina, E.~Oset and B.~S.~Zou, Phys.\ Rev.\ Lett.\  {\bf 105} (2010)
232001.
\bibitem{Wujj2} J.~J.~Wu, R.~Molina, E.~Oset and B.~S.~Zou, Phys.\ Rev.\ C {\bf 84} (2011)
015202.

\bibitem{Wangwl} W. L. Wang, F. Huang, Z. Y. Zhang, B. S. Zou, Phys. Rev. C {\bf 84} (2011)
015203.

\bibitem{Wujj-Lee} J. J. Wu, T. S. H. Lee, B. S. Zou, Phys. Rev. C {\bf 85} (2012)
044002.

\bibitem{Liuxiang} Z.~C.~Yang, Z.~F.~Sun, J.~He, X.~Liu, S.~L.~Zhu,
Chin.\ Phys.\ C {\bf 36} (2012) 6.

\bibitem{Yuansg} S.~G.~Yuan, K.~W.~Wei, J.~He, H.~S.~Xu, B.~S.~Zou,
Eur.\ Phys.\ J.\ A {\bf 48} (2012) 61.

\bibitem{Gobbi} C. Gobbi, D. O. Riska, N. N. Scoccola, Phys.\ Lett.\  B {\bf 296}
(1992) 166.

\bibitem{Hofmann} J. Hofmann, M. F. M. Lutz, Nucl.\ Phys.\  A {\bf 763}
(2005) 90.

\bibitem{Wujj3} J.~J.~Wu and B.~S.~Zou, Phys.\ Lett.\ B {\bf 709} (2012)
70.
\bibitem{Zb} A.~Bondar {\it et al.}  [Belle Collaboration], Phys.\ Rev.\ Lett.\  {\bf 108} (2012)
122001.
\end{thebibliography}



\end{document}